\documentclass{ifacconf}

\usepackage{lipsum}
\usepackage{mwe}
\usepackage{graphicx}
\usepackage{amsmath}
\usepackage{algorithm}
\usepackage{algpseudocode}
\usepackage{natbib}        
\usepackage{tikz}

\usepackage{pgfplots}
\usepgfplotslibrary{fillbetween}
\pgfplotsset{compat=newest} 
\pgfplotsset{plot coordinates/math parser=false}
\usepackage{xcolor}
\usepackage{enumitem}
\newcommand{\JS}[1]{{\color{blue} #1}}

\newcommand{\plotLine}{0.75pt}
\newcommand{\plotDotted}{1.25pt}
\newcommand{\plotBox}{1pt}

\newcommand{\plotArrow}{1pt}

\definecolor{color0}{rgb}{0.8235,0,0} 
\definecolor{color1}{rgb}{0.07843,0.549,0.07843} 
\definecolor{color2}{rgb}{0,0,1} 
\definecolor{color6}{rgb}{0.3451,0.3451,0.3451} 
\definecolor{color8}{rgb}{0,0,0} 
\begin{document}
\begin{frontmatter}

\title{\vspace{-10pt}Data-Driven Track Following Control for Dual Stage-Actuator Hard Disk Drives}

\author[First]{Nikhil Potu Surya Prakash} 
\author[First]{Joohwan Seo}
\author[Second]{Alexander Rose} 
\author[First]{Roberto Horowitz}

\address[First]{University of California, Berkeley, Berkeley, CA, 94720 USA 
}
\address[Second]{Leibniz University Hannover, Germany, 
   Hanover, Germany\\
e-mail: \{nikhilps, joohwan\_seo, rose1993, horowitz\}@berkeley.edu}

\begin{abstract}                
In this paper, we present a frequency domain data-driven feedback control design methodology for the design of tracking controllers for hard disk drives with two-stage actuator as a part of the open invited track 'Benchmark Problem on Control System Design of Hard Disk Drive with a
Dual-Stage Actuator' in the IFAC World Congress 2023 (Yokohoma, Japan). The benchmark models are    
%
Compared to the traditional controller design, we improve robustness and avoid model mismatch by using multiple frequency response plant measurements directly instead of plant models.
%
%
Disturbance rejection and corresponding error minimization is posed as an $H_2$ norm minimization problem with $H_\infty$ and $H_2$ norm constraints.
$H_\infty$ norm constraints are used to shape the closed loop transfer functions and ensure closed loop stability and $H_2$ norm constraints are used to constrain and/or minimize the variance of relevant.
\end{abstract}


\end{frontmatter}

\section{Introduction}\label{sec:intro}
%
Currently, two main types of drives are used for data storage: Hard Disk Drives (HDD) and Solid State Drives (SSD).
While SSDs are advantageous for use in Personal Computers (PC) due to high read/write speed and size, HDDs are dominant for data storage devices in data centres due to their high robustness and comparatively low cost (\cite{ChPrHo:22}). 
Cost and performance of a HDD are strongly connected to the storage capacity, which highly depends on the spacing between successive tracks on which the data is stored.

This spacing is limited by the sensitivity of the controller to external disturbances, measurement noises and track runout. 
The major sources of disturbance in HDDs are the fan-induced vibrations from the cooling systems and the rotational vibrations from other HDDs in a storage box at the data centers.
These disturbances have a wide bandwidth with high-frequency components up to 10 kHz.

Thereby, low-frequency disturbance compensation is done by a Voice Coil Motor (VCM) with higher strokes and high-frequency disturbance compensation is realized by a low stroke Piezo-electric actuators (PZT). Both these actuators together form the Dual Stage Actuation (DSA) system of an HDD.
PZTs can commonly reach one to three tracks.

Furthermore, due to manufacturing tolerances and due to the dependence on the temperature, the plants of each HDD are different.
%
Therefore, robust and precise controllers are needed to handle the disturbances for variations of plants.
Data-driven feedback controllers are capable of handling this complex task.

Data-driven feedback control design approaches from frequency response measurements [\cite{GaKaLo:10,KaKa:17,KaNiZh:16}] for dual stage systems have been developed in \cite{Ba:17, BaShHo:18} to suppress the disturbances. 
Controllers for disturbance rejection and data-driven $H_\infty$ synthesis controllers for triple-stage actuator systems have also been developed in \cite{Ni:22}. 
Though these controllers are robust to the variations in the plant models, a common controller might not be the optimal controller for each individual HDD as there might be variations in the plants and the disturbances. 
To address these issues, data driven feedforward controllers have been developed in \cite{ShHo:20,ShHo:19,ShHo:21} based on the frequency response measurement of disturbance processes on top of the robust feedforward controllers and another add-on adaptive feedforward controller has been developed in \cite{ChPrHo:22} on top of 
the robust feedback controller to account for plant variations. 

%
%
%
%
In this paper, we present a data-driven mixed $H_2/H_\infty$ synthesis framework for designing robust feedback controllers for a DSA HDD with a VCM and a PZT actuator. The design is performed on the benchmark HDD actuator models from \cite{At:20} as a art of the open invited track session 'Benchmark Problem on Control System Design of Hard Disk Drive with a
Dual-Stage Actuator' at IFAC World Congress 2023 to be held in Yokohoma, Japan.
For this purpose we use multiple plant measurements for a benchmark DSA HDD.
The disturbance rejection controller design will be posed as a convex optimization problem with the objective to minimize the average $H_2$ norm of the output due to various process noises subject to constraints on $H_\infty$ and $H_2$ norms of appropriate closed loop transfer functions.
The $H_\infty$ constraints will be used to shape the closed loop transfer functions using numerically shaped weights in the frequency domain and guarantee stability of the closed loop system (\cite{Ba:17}). 
By the $H_2$ norm constraints the variances of various signals of interest are treated.

In section~\ref{sec:Preliminaries}, we introduce the plant of the DSA HDD as block diagram, give the controller factorization with the frequency response of the VCM \& PZT, define the closed-loop transfer functions and show the frequency responses of the disturbance.
Section~\ref{sec:controller} describes our data-driven control design using $H_2$- and $H_\infty$-norm.
For both, we define the constraints.
Following from the $H_2$-objective, the minimum $H_2$-norm controller is defined.
In section~\ref{sec:Results}, we show the simulation results of the controller for nine different plants.
\section{Preliminaries}\label{sec:Preliminaries}
%
In this paper, we use the discrete time domain (z-domain).
Note that the generalization to the case of continuous time (s-domain) is straightforward.
The DSA HDD plant model $G(z)$ 
is a Multiple Input Single Output (MISO) system with two inputs and determines the displacement of the head from the output of the two controllers.
%
%
%
In the following we use the shorthand notation with the dependence on $z$.
We use a set of frequency responses $G\left(e^{j\omega T_s}\right)$, with $\omega\!\in\!\Omega\!=\!(-\frac{\pi}{T_s},\frac{\pi}{T_s})$ and sampling time $T_s$ to characterize the plant.

\subsection{Control structure of DSA HDDs}

\begin{figure}[h!]
\begin{center}
\includegraphics[width=8.4 cm]{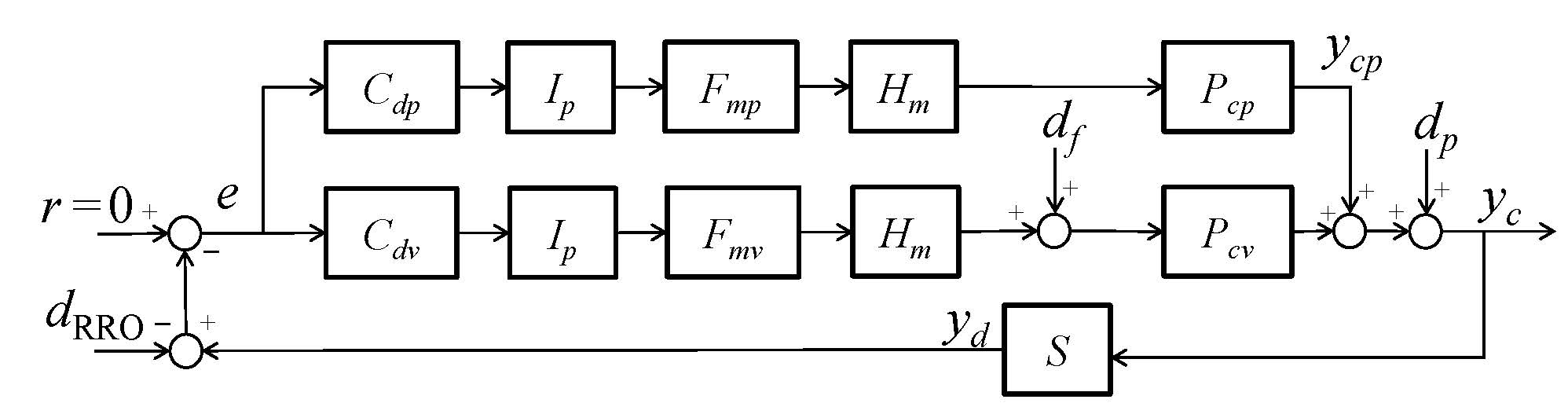}
\includegraphics[width=8.4 cm]{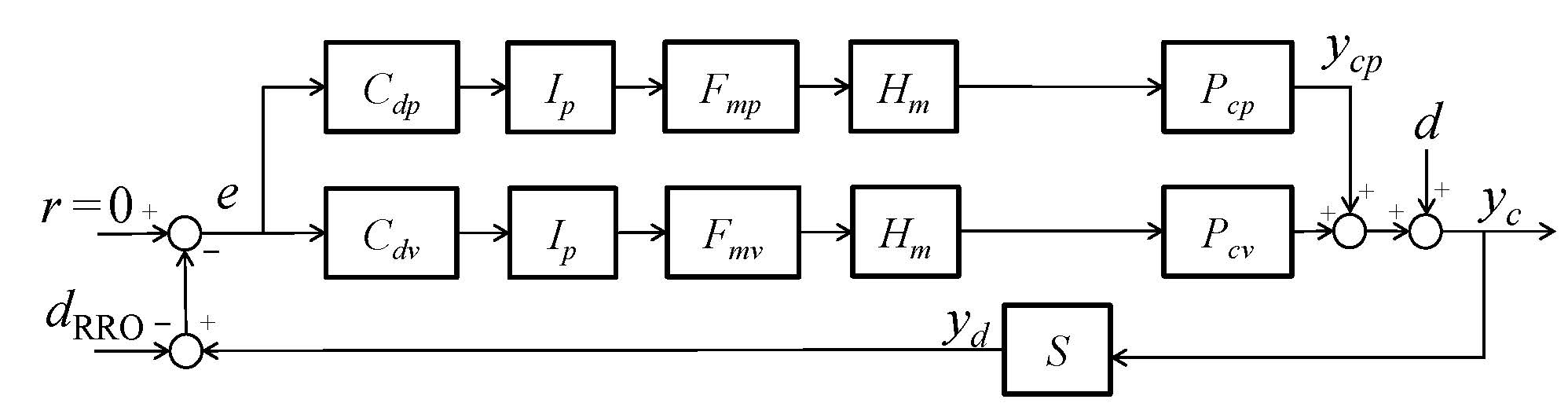}
\vspace{-5pt}
\caption{Control block diagram for a Dual Stage Actuator HDD (top) and control block diagram with lumped disturbance (bottom).} 
\label{fig:BDDSA}
\end{center}
\end{figure}
\begin{figure}[h!]
\centering
\input{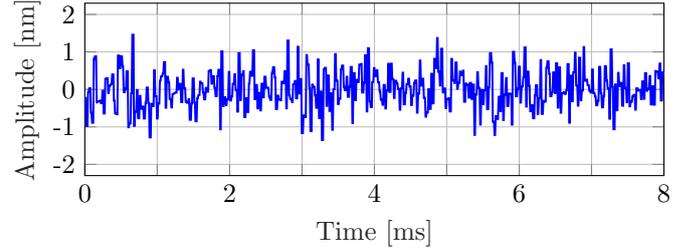}
\vspace{-20pt}
\caption{Repeatable runout used for the simulation.}
\label{fig:dRRO}
\end{figure}

The block diagram of the DSA is shown in Fig.~\ref{fig:BDDSA}.
$r$ is the reference that needs to be tracked and 
$r$ is set to zero for track following purposes.
The position error signal $e$ results from the reference, the sampled total head position $y_d$ and the repeatable runout $d_{RRO}$ (see Fig.~\ref{fig:dRRO}), which is the track offset from the ideal track.
This error feeds into the discrete time controllers $C$ for the actuators of the two plants, the VCM plant and the PZT plant, of the MIMO system.
The path for PZT has subscript $dp$ and the path for VCM has subscript $dv$.

\begin{figure}[h!]
\centering
\input{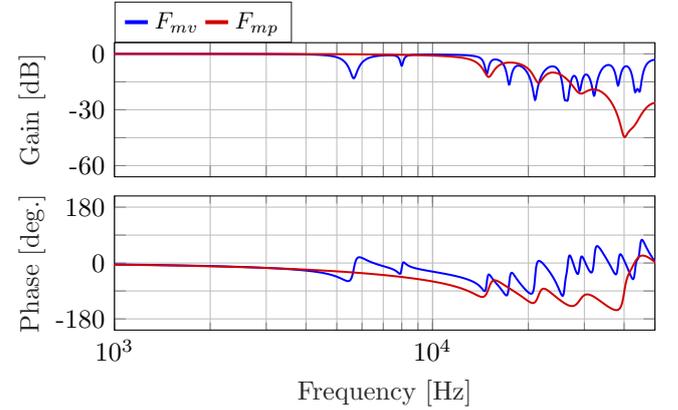}
\vspace{-10pt}
\caption{Multi-rate filters for VCM (blue) and PZT (red) actuators}
\label{fig:MRF}
\end{figure}

With the interpolation block $I_p$ the output signal of the controller is sampled up to a higher frequency for the multi-rate filters $F_{mp}$ and $F_{mv}$ for PZT and VCM respectively (see Fig.\ref{fig:MRF}).
The zero order hold (ZOH) process $H_m$ converts the discrete signal into a continuous signal.
For both the paths, same $H_m$ is applied.
While the rotational vibrations $d_f$ caused by other HDDs in the storage box only affect the VCM path, the disturbances from the fans of the cooling system $d_p$ act on both.
For the design, we assume that $d_f$ and $d_p$ are generated by zero mean white noise of unit variance $\bar{d}_f$ and $\bar{d}_p$, and filtered by transfer function $D_f$ and $D_p$, respectively.
Both disturbances can be combined to $d \!=\! d_p \!+\! P_{cv}d_f$ using block diagram manipulations.
$y_{cp}$ denotes the displacement of the PZT actuator which is of special interest as it can only have a finite stroke to stay within the linear regime.
The total head position $y_c$ is sampled with $S$ to the discrete time signal $y_d$.




\begin{figure}[h!]
\centering
\input{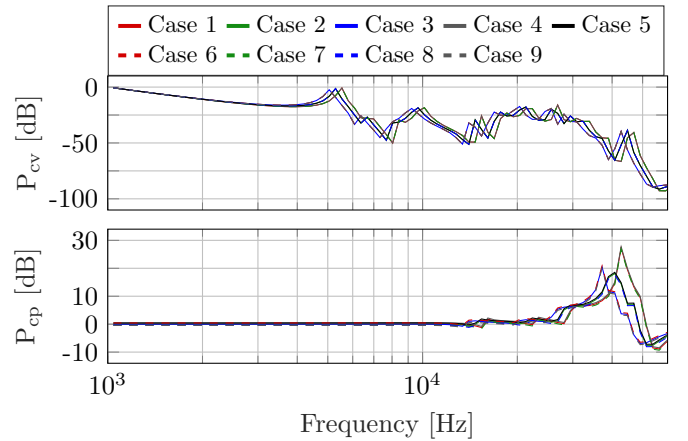}
\vspace{-10pt}
\caption{ Frequency responses of the VCM plant (top) and PZT plant (bottom) }
\label{fig:VCMPZT}
\end{figure}
\begin{figure}[h!]
\centering
\input{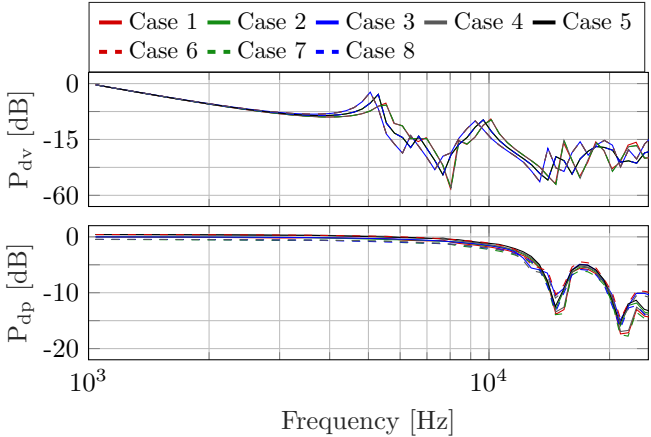}
\vspace{-10pt}
\caption{ Frequency responses from input of $F_{mv}$ to output of $P_{cv}$ (top) and input of $F_{mp}$ to output of $P_{cp}$ (bottom) }
\label{fig:FVCMPZT}
\end{figure}
The magnitude frequency responses of the benchmark model for the VCM and
PZT actuators are shown in Fig. \ref{fig:VCMPZT}. Some typical resonance frequency variations are also shown. The combined responses with the multi-rate filters are also presented in Fig. \ref{fig:FVCMPZT}.
The variability in the resonance frequencies usually limit the attainable bandwidth of the servo system. 
Therefore, piggy-back PZT actuators, which can only achieve a stroke of a few tracks but have higher resonance modes than the VCM, are used to increase the overall disturbance attenuation of the servo system. 
In the model used for our case study, the PZT actuator has a resonance mode with a nominal resonance frequency of 40 kHz. 


In the production of HDDs, manufacturing tolerances lead to slightly different plants of the DSA systems.
Therefore, the dynamics of the actuators and also the frequency responses differ between the drives, see Fig.\ref{fig:VCMPZT}.
Moreover, even for a specific actuator, the dynamics may vary due to environmental variations such as temperature.

\begin{figure}[h!]
\centering
\input{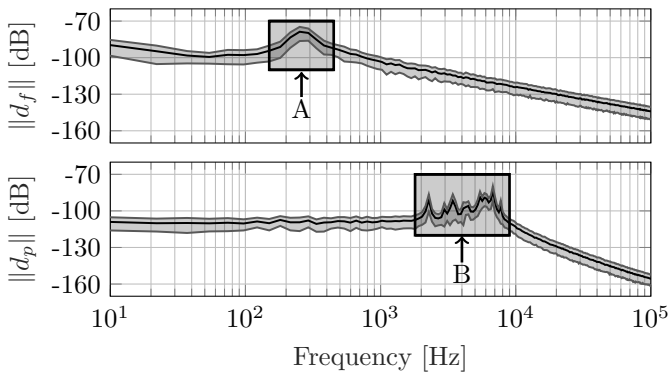}
\vspace{-20pt}
\caption{Typical frequency spectrum of the disturbances $d_f$ (top) and $d_p$ (bottom) used in Figs. \ref{fig:BDDSA}. Boxes A and B show the resonant components}
\label{fig:dfdp}
\end{figure}

In large data centers, millions of enterprise HDDs are stacked in the server boxes. If several HDDs are placed next to each other, vibrations due to neighboring drives or cooling fans can affect the hard disks differently.
%
It is common in industry to incorporate the effects of track run-out, windage, internal, and external disturbances in a disturbance filter $D(z)$.
A typical example of a frequency response of $D(z)$ is given in Fig.~\ref{fig:dfdp}.
%
Since the influence of $D(z)$ on the position error $e$ is much larger than measurement noise $n$ and runout $d_{RRO}$, we focus on $D(z)$. However, we note that 
$n$ and $d_{RRO}$ can be easily incorporated into our framework.

%

\subsection{Controller Factorization}
The vector controller transfer function can be factorized as $K \!=\! XY^{-1}$ following the convention from \cite{KaKa:17} as
\begin{equation}
\begin{aligned}
    X(z) &= \mathcal{X}_pz^p+\mathcal{X}_{p-1}z^{p-1}+\dots+\mathcal{X}_0, \nonumber \\
    Y(z) &= z^p+\mathcal{Y}_{p-1}z^{p-1}+\dots+\mathcal{Y}_0,
\end{aligned}
\end{equation}
where $p$ is the controller order and the controller parameters are $\{\mathcal{Y}_{p-1},\dots, \mathcal{Y}_0 \} \!\in\! \mathcal{R}$ and $\{\mathcal{X}_p,\mathcal{X}_{p-1},\dots, \mathcal{X}_0 \} \!\in\! \mathcal{R}^{n\times 1}$. For the remainder of the paper, the convention $X_i = \mathcal{X}_p(i)z^p+\mathcal{X}_{p-1}(i)z^{p-1}+\dots+\mathcal{X}_0(i)$ will be used to denote the $i^{th}$ row of $X(z)$.

With the controller factorization presented above, various closed loop transfer functions can be calculated as follows
\begin{equation}\label{eq:SensitivityCF}
\begin{aligned}
    &S_{d \rightarrow e}  &= \frac{Y}{Y+GX} , \; 
    &U_{d \rightarrow u_{VCM}} &= \frac{X_1}{Y+GX}, \\
    &U_{d \rightarrow u_{PZT}} &= \frac{X_2}{Y+GX}, \;  
    &Y_{d \rightarrow y_{PZT}} &= \frac{G_{PZT}X_2}{Y+GX}
\end{aligned}
\end{equation}
where $G = [P_{cv} \; P_{cp} ]$ is the DSA plant.  
\section{Data-Driven Control Design} \label{sec:controller}
In this section, various objectives and constraints that will be used in the disturbance rejection controller design will be formulated in such a way that a convex optimization solver can obtain a solution. It's important to pose the problem as a convex optimization problem as some of the constraints involve matrix inequalities and any nonlinearities in the matrix inequalities cannot be handled by most of the existing optimization solvers. The conversion to a convex optimization problem is an attractive method
for robust controller design as $H_2$ constraints and objectives can also be handled by the framework. The formulation of the constraints and objectives will be demonstrated for one of the closed loop transfer functions, but the same procedure can be followed for any of the closed loop transfer functions. 

The $H_\infty$ constraints are used to ensure stability and shape the closed loop transfer functions and $H_2$ constraints are used to constrain the variance of signals of interest. In the data-driven design, the constraints will only be enforced at the frequencies at which the responses are available. It is assumed that the data set is rich enough i.e., the frequencies at which the data is collected represent the key characteristics of the plants. Otherwise there is a possibility of obtaining controllers that might not stabilize the plant. Some such cases are 
\begin{enumerate}[label=(\roman*)]
    \item when the controller order is comparable to the number of data points, the obtained solution might be an overfit to the data and the real system might not be stabilized by the controller,
    \item when the key characteristics like the resonant peaks are not captured by the data, again the obtained solution might not stabilize the actual plant.
\end{enumerate}

\subsection{Constraints and Objectives}
\subsubsection{$H_\infty$ Constraints:}
$H_\infty$ constraints are typically used to shape various closed loop transfer functions with appropriate weighting transfer functions. In the data driven control framework, a feasible controller that satisfies the $H_\infty$ constraints is guaranteed to stabilize the closed loop system [\cite{Ba:17}]. A typical $H_\infty$ constraint on the weighted sensitivity of the DSA from the disturbance $d$ to the position error signal $e$ can be formulated as $||W(\omega)S_{d \rightarrow e}(e^{j\omega})||_\infty \!\leq\! \gamma$, where $W(\omega)$ is a numerically shaped weight for all the frequencies $\omega\!\in\!\Omega \!=\! (-\frac{\pi}{T_s},\frac{\pi}{T_s})$.

Using the definition for sensitivity in (\ref{eq:SensitivityCF}) with the controller factorization, this constraint can be equivalently written as (Ch 5 of \cite{Ba:17})
\begin{equation}\label{eq:HinfConstraint1}
    \gamma^{-1}|W(\omega)Y(e^{j\omega})| < Re[Y(e^{j\omega})+G(\omega)X(e^{j\omega})].
\end{equation}
Note that for a MISO system, this constraint is a convex constraint as it can be expressed as a Second Order Cone constraint. Any of the available Second Order Conic Program (SOCP) solvers can handle these constraints. For practical purposes, as strict inequality constraints cannot be handled by optimization solvers, the strict inequalities are replaced with slack inequalities with the help of a small tolerance. 
The $H_\infty$ constraints that are used in the current design are as follows:
\begin{subequations}\label{eq:VCMHinf}\begin{align}
&\mbox{Single Stage (VCM):}\nonumber \\
&\begin{aligned}
||W_{S^{VCM}_{d \rightarrow e}}S^{VCM,i}_{d \rightarrow e}||_\infty &< 1, \\
||W_{U^{VCM}_{d \rightarrow u_v}}U^{VCM,i}_{d \rightarrow u_v}||_\infty &< 1, 
\end{aligned}\\[10pt]
&\mbox{Dual Stage (DSA):}\nonumber \\
&\begin{aligned}
||W_{S^{DSA}_{d \rightarrow e}}S^{DSA,i}_{d \rightarrow e}||_\infty &< 1, \\
||W_{U^{DSA}_{d \rightarrow u_v}}U^{DSA,i}_{d \rightarrow u_v}||_\infty &< 1.
\end{aligned}
\end{align}\end{subequations}
Here the superscript VCM refers to the case when only the VCM is operating i.e., when $G = [P_{cv} \; 0]$ and the superscript DSA refers to the case when both the VCM and PZT are operating i.e., when $G = [P_{cv} \; P_{cp}]$. The superscript $i$ is used to represent the $i^{th}$ plant i.e., the plant's frequency response $G$ would be of the $i^{th}$ set of VCM and PZT plants.

The first set of constraints on just the single stage are imposed to ensure that the system remains stable even PZT actuator fails. The weights on the VCM's input for the both the stages are designed in such a way that the resonant modes of the VCM actuator are 
not excited.
$H_\infty$ constraints on the PZT's inputs are not imposed as the multi-rate filters are already designed in a way that the PZT's resonant modes are not excited. But, if multi-rate filters are not used, $H_\infty$  constraints can also be introduced on the PZT's inputs. The inverse weights used for the design are shown with dashed blue lines in Figs. \ref{fig:Sgroup} and \ref{fig:Ugroup}.

\subsubsection{$H_2$ Constraints:}
$H_2$ constraints are used to constrain the variances of various signals. Parseval's theorem states that the time domain and frequency domain $L_2$ norms of a signal are equivalent. This means that the $H_2$ norm of a transfer function is equal to the $L_2$ norm of the transfer function's impulse response (Note that $H_2$ norm is used for a transfer function whereas $L_2$ norm is used for a signal).

The convolution of the impulse response with the input signal gives us the system's response to the input signal. For a normally distributed random variable, the confidence of finding the random variable within three standard deviations from the mean is $99.7\%$. Though hard limits on the magnitudes of the signals cannot be enforced in the frequency domain design unlike in state space techniques like Model Predictive Control (MPC), we can constrain the signal's variance to increase the confidence that the signal does not go beyond the limit $99.7 \%$ of the time. This can be done by restricting the $L_2$ norm of the zero mean signal to be always less than a third of the limit. 

The constraints can be applied on various signals like the inputs to the actuators, the stroke limits of the actuators, the overall position error signal etc. In case of multiple plants, we can either constrain the $L_2$ norms of the signals for each of the plants or constrain the average $L_2$ norm  of the signals for all the plants. A PZT actuator's displacement needs to be restricted for it to behave linearly. This forms a classic example of an $H_2$ norm constraint. The constraint in this case can be formulated as constraining the average squared $H_2$ norm of the output of the PZT due to various process noises. This would constrain the average variance of the output of the PZT actuator. The maximum value of the variance can be set based on the confidence with which we want the output of the PZT to stay within the desired limits. Typically the stroke of a PZT is set to be around one to three tracks. For the current design, the average stroke is limited to be $\mu = 50 nm$\JS{.}
This constraint can be formulated using Parseval's theorem as

\begin{equation}\label{eq:H2Constraints1}
\begin{aligned}
    ||y^i_{cp}||^2_2 =\, &||Y^i_{d \rightarrow y_{cp}}D_p||^2_2\\
    +&||Y^i_{d \rightarrow y_{cp}}P^i_{dv}D_f||^2_2\\
    \implies\! \frac{1}{l}\sum_{i=1}^{l}||y^i_{cp}||^2_2 =\, & \frac{1}{l}\sum_{i=1}^{l}(||Y^i_{d \rightarrow y_{cp}}D_p||^2_2\\
    &+||Y^i_{d \rightarrow y_{cp}}P^i_{dv}D_f||^2_2)\leq \mu, 
\end{aligned}
\end{equation}
where the subscript $i$ corresponds to the $i^{th}$ plant.
The $H_2$ norm for this PZT output transfer function is given by
\begin{equation}\label{eq:H2Constraints2}
    \hspace{-5pt}||Y^i_{d \rightarrow y_{cp}}D_p||^2_2 \!=\! \frac{T_s}{2\pi} \!\int\limits_{\Omega} \!tr[(Y^i_{d \rightarrow y_{cp}}D_p)^*Y^i_{d \rightarrow y_{cp}}D_p]d\omega.
\end{equation}
where $*$ denotes the complex conjugate.

Using slack variables $\Gamma^{i}_{y_{cp}}$ and $\Lambda^{i}_{y_{cp}}$, the $H_2$ constraint in Eq. (\ref{eq:H2Constraints1}) can be reformulated into two constraints as,
\begin{subequations}\begin{align}
    \label{eq:H2Constraints3}
     &\frac{1}{l}\sum_{i=1}^{l}\frac{T_s}{2\pi} \int\limits_{\Omega}\! tr[\Gamma^{i}_{y_{cp}}(\omega)+\Lambda^{i}_{y_{cp}}(\omega)]d\omega \leq \mu,\\
    &\begin{aligned}\label{eq:H2Constraints4}
        &(Y^i_{d \rightarrow y_{cp}} D_p)^*  Y^i_{d \rightarrow y_{cp}}D_p \preceq \Gamma^{i}_{y_{cp}}.  \\
    \end{aligned} \\
        &\begin{aligned}\label{eq:H2Constraints4b}
        &(Y^i_{d \rightarrow y_{cp}}P^i_{dv} D_f)^*   Y^i_{d \rightarrow y_{cp}}P^i_{dv}D_p \preceq \Lambda^{i}_{y_{cp}}.  \\
    \end{aligned}
\end{align}\end{subequations}
Since the responses are available only at a finite set of frequencies, the integral in Eq.~(\ref{eq:H2Constraints3}) will be evaluated approximately using a trapezoidal rule by replacing the function $\Gamma^{i}_{y_{cp}}(\omega)$ with a finite set of $\Gamma^{k,i}_{y_{cp}}$ for $k\!=\!1\mbox{ to } N$ where $N$ is the number of frequency responses available per plant. The matrix inequalities in Eq.~(\ref{eq:H2Constraints4}) \& (\ref{eq:H2Constraints4b}) can be reformulated using controller factorization and the Schur complement lemma as
\begin{subequations}
\begin{align}
&\begin{aligned}\label{eq:H2Constraints5a}
\begin{bmatrix}
        \Gamma^{k,i}_{y_{cp}} & X_1D_p \\
        (X_1D_p)^* & (Y+G^iX)^*(Y+G^iX) 
\end{bmatrix}(\omega_k) &\succeq 0.\\
\end{aligned}\\
&\begin{aligned}\label{eq:H2Constraints5b}
\begin{bmatrix}
        \Lambda^{k,i}_{y_{cp}} & X_1P^i_{dv}D_p \\
        (X_1P^i_{dv}D_p)^* & (Y+G^iX)^*(Y+G^iX) 
\end{bmatrix}(\omega_k) &\succeq 0.\\
\end{aligned}
\end{align}
\end{subequations}

Notice that this  is a non convex constraint, as ${P^i}^*P^i \!=\! (Y\!+\!G^iX)^*(Y\!+\!G^iX)$ is quadratic. However, this quadratic part can be linearized using a first order Taylor series expansion around a nominal controller $K_c \!=\! X_c/Y_c$, as shown in \cite{{KaKa:17}}.
\begin{equation}
    (P^i)^*P \approx {P^i_c}^*P^i_c+(\Delta P^i)^*P^i_c+{P^i_c}^*(\Delta P^i),
\end{equation}
where $P^i_c = Y_c+G^iX_c$ and $\Delta P^i = P^i-P^i_c$
This leads to
\begin{subequations}
\begin{align}
&\begin{aligned}
\label{eq:H2Constraints7a}
\begin{bmatrix}
    \Gamma^{k,i}_{y_{cp}} & X_1D_p \\
    (X_1D_p)^* & (P^i_c)^*P^i_c\!+\!(\Delta P^i)^*P^i_c\!+\!{P^i_c}^*(\Delta P^i) 
\end{bmatrix}\!(\omega_k) \!\succeq\! 0.
\end{aligned}\\
&\begin{aligned}
\label{eq:H2Constraints7b}
\begin{bmatrix}
    \Lambda^{k,i}_{y_{cp}} & X_1P^i_{dv}D_p \\
    (X_1P^i_{dv}D_p)^* & (P^i_c)^*P^i_c\!+\!(\Delta P^i)^*P^i_c\!+\!{P^i_c}^*(\Delta P^i) 
\end{bmatrix}\!(\omega_k) \!\succeq\! 0.
\end{aligned}
\end{align}
\end{subequations}

\subsubsection{$H_2$ Objective:} $H_2$ objectives can be used in cases when the variances of signals need to be minimized. Hence $H_2$ objective becomes an ideal choice for disturbance rejection problems. The objective can be formulated to minimize the variance of the tracking error i.e., the position error signal $e$ due to various process noises. The position error signal due to the disturbance $\bar{d}_f$ and $\bar{d}_p$ is given by  $S_{d \rightarrow e} D\bar{d}$.


The optimization problem in it's raw form can be written as
\begin{equation}\label{eq:H2Obj1}
\begin{aligned}
& \underset{X,Y}{\text{min}}
& & \frac{1}{l}\sum_{i=1}^{l}(||S^{DSA,i}_{d \rightarrow e}D_p||^2_2+||S^{DSA,i}_{d \rightarrow e}P^i_{dv}D_f||^2_2) \\[5pt]
& \text{s.t.}
& & H_\infty \mbox{ and } H_2 \mbox{ constraints}. 
\end{aligned}
\end{equation}

Using a similar approach shown for the $H_2$ constraints, we can formulate the optimization problem using slack variables $\Gamma^i_{S_{d \rightarrow e}}$ and $\Lambda^i_{S_{d \rightarrow e}}$ as follows
\begin{equation}\label{eq:H2Obj2}
\begin{aligned}
& \underset{X,Y}{\text{min}}
& &  \frac{1}{l}\sum_{i=1}^{l}\int_{\Omega} tr[\Gamma^i_{S_{d \rightarrow e}}(\omega)+\Lambda^i_{S_{d \rightarrow e}}(\omega)]d\omega \\[5pt]
& \text{s.t.}
& & \begin{bmatrix}
        \Gamma^i_{S_{d \rightarrow e}}(\omega) & YD_p \\
        (YD_p)^* & (Y+G^iX)^*(Y+G^iX) 
\end{bmatrix} \succeq 0, \\
& & & \begin{bmatrix}
        \Lambda^i_{S_{d \rightarrow e}}(\omega) & YP^i_{dv}D_p \\
        (YP^i_{dv}D_p)^* & (Y+G^iX)^*(Y+G^iX) 
\end{bmatrix} \succeq 0, \\
& & & H_\infty \mbox{ and } H_2 \mbox{ constraints}. 
\end{aligned}
\end{equation}

\subsection{Minimum $H_2$ norm controller}
As was shown previously, the $H_2$ norm minimization is a non-convex problem. The quadratic part can be linearized around a nominal controller to make the constraint convex or more precisely a Linear matrix Inequality (LMI). The final controller can be designed by first obtaining the nominal  controller using just $H_\infty$ constraints without any objective and then linearizing around it to obtain a better controller with the $H_2$ constraints and $H_2$ objective. The detailed algorithm is shown algorithm \ref{alg:H2}.

\begin{algorithm}
\caption{Iterative algorithm for minimum $H_2$ norm controller}\label{alg:H2}
\begin{algorithmic}[1]
\Require $N_{iter} \in \mathcal{Z^+}$ \Comment{number of iterations required for satisfactory convergence}
\State Define $H_\infty$ Constraints
\State Optimize($H_\infty$ Constraints)
\State Obtain initial controller $K_0 \!=\! X_0/Y_0$
\State Define linearized $H_2$ Constraints with $X_0,Y_0$
\While{$k\leq N_{iter} $}
    \State Optimize($H_2$ Objective, $H_\infty , H_2$ Constraints)
    \State Obtain $K_k \!=\! X_k/Y_k$
    \State Define linearized $H_2$ Constraints with $X_k,Y_k$
\EndWhile
\end{algorithmic}
\end{algorithm}

\section{Results}\label{sec:Results}
The optimization problem with all the constraints presented in the previous section was set up using MOSEK and YALMIP in MATLAB R2021a. The optimization problem was solved on a Windows 11 Lenovo Thinkpad PC with AMD Ryzen 7 4700U 2.00 GHz processor once with 300 equally spaced frequency data points and once with 10000 equally spaced frequency data points  from 0 to 25200 Hz. The design process took on average 1040 seconds when the data points were 300 and 7320 seconds for 10000 data points when the controller order was 12 and the number of iterations used for the minimization of $H_2$ norm for both the cases were 2. The obtained controller was able to stabilize the closed loop system.

\begin{figure}[h!]
\centering
\input{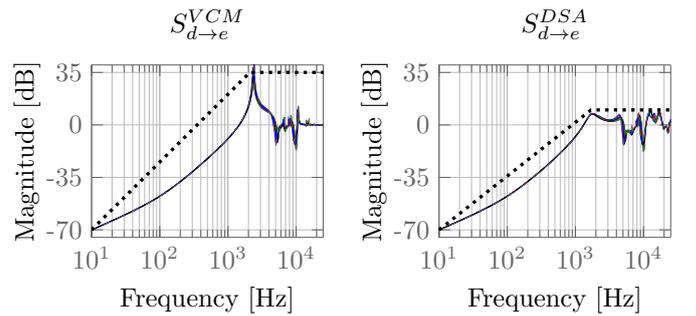}
\vspace{-20pt}
\caption{Bode plots of closed loop disturbance to error sensitivities for the single stage $S^{VCM}_{d \rightarrow e}$ (left) and the dual stage $S^{DSA}_{d \rightarrow e}$ (right) for all the nine cases. Dotted blue lines show their corresponding inverse weights.}
\label{fig:Sgroup}
\end{figure}

\begin{figure}[h!]
\centering
\vspace{-20pt}
\input{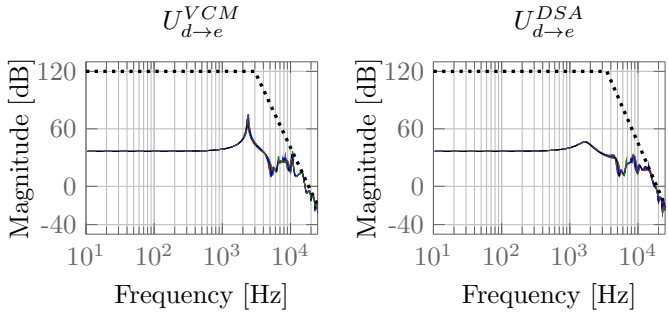}
\vspace{-20pt}
\caption{Bode plots of closed loop disturbance to VCM input transfer functions for the single stage $U^{VCM}_{d \rightarrow u_v}$ (left) and dual stage $U^{DSA}_{d \rightarrow u_v}$ (right) for all the nine cases. Dotted blue lines show their corresponding inverse weights.}
\label{fig:Ugroup}
\end{figure}

The closed loop disturbance to error sensitivity and the disturbance to input of VCM along with the corresponding inverse weights (dotted lines) used in the $H_\infty$ constraints are shown in Fig. \ref{fig:Sgroup} and Fig. \ref{fig:Ugroup} for both the single stage and the dual stage systems. The plots show that the $H_\infty$ constraints are respected. It can also be seen that the low frequency disturbance attenuation characteristics of the designed controller are suitable to meet our design requirements. A bandwidth of 1 kHz was achieved for the closed loop system. 

\begin{figure}[h!]
\centering
\input{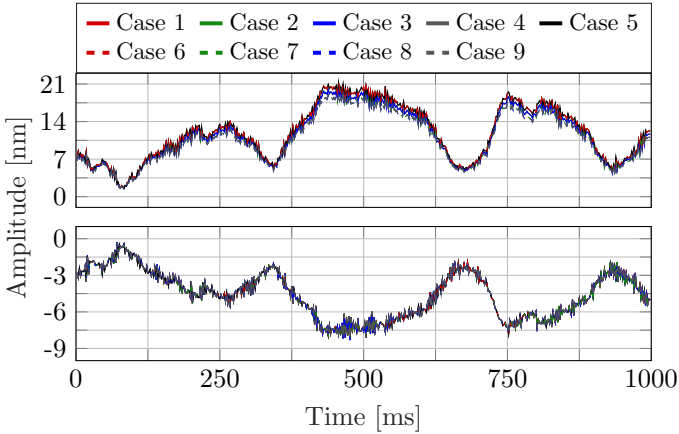}
\vspace{-20pt}
\caption{Moving means of total head position of the DSA system (bottom) and the output of the PZT actuator (top) in nanometers for all the 9 cases.}
\label{fig:ycycp}
\end{figure}

\begin{figure}[h!]
\centering
\vspace{-10pt}
%
%
\begin{tikzpicture}

\begin{axis}[%
width=1.2in,
height=0.9in,
at={(0in,0in)},
scale only axis,
xmin=0.4,
xmax=9.6,
xlabel style={font=\color{white!15!black}},
xlabel={Case number},
xtick={1,2,3,4,5,6,7,8,9},
xticklabels={1,,3,,5,,7,,9},
ymin=23.8,
ymax=28.2,
ylabel style={font=\color{white!15!black}},
ylabel={Value [nm]},
y label style={at={(-0.16, 0.5)}},
ytick={24,25,26,27,28},
axis background/.style={fill=white},
title style={font=\bfseries},
title={Max of $|y_{cp}|$},
xmajorgrids,
ymajorgrids
]
\addplot [only marks, mark=triangle*]
table[row sep=crcr]{%
1	26.7650283106618\\
2	25.6457836828652\\
3	25.9119141163557\\
4	27.887167404703\\
5	26.8320481623992\\
6	27.0423812373186\\
7	25.630673761021\\
8	24.4465518326117\\
9	24.7748680121751\\
};
\end{axis}

\begin{axis}[%
width=1.2in,
height=0.9in,
at={(1.8in,0in)},
scale only axis,
xmin=0.6,
xmax=9.4,
xlabel style={font=\color{white!15!black}},
xlabel={Case number},
xtick={1,2,3,4,5,6,7,8,9},
xticklabels={1,,3,,5,,7,,9},
ymin=16.9,
ymax=19.1,
ylabel style={font=\color{white!15!black}},
ylabel={Value [\% Track width]},
y label style={at={(-0.14, 0.5)}},
ytick={17,17.5,18,18.5,19},
yticklabels={17,,18,,19},
axis background/.style={fill=white},
title style={font=\bfseries},
title={$3\sigma$ of $y_{c}$},
xmajorgrids,
ymajorgrids
]
\addplot [only marks, mark=triangle*]
  table[row sep=crcr]{%
1	17.6082646278462\\
2	17.6547703171586\\
3	18.2849688179229\\
4	17.3114323341864\\
5	17.3459362585362\\
6	17.939147125464\\
7	17.9420770987013\\
8	17.9996699743295\\
9	18.6684711140404\\
};
\end{axis}

\end{tikzpicture}%
\vspace{-10pt}
\caption{Maximum value of the output of PZT in each case in nm (left) and $3\sigma$ values of the total head position as a percentage of track width in each case (right) }
\label{fig:maxycp3sigmayc}
\end{figure}
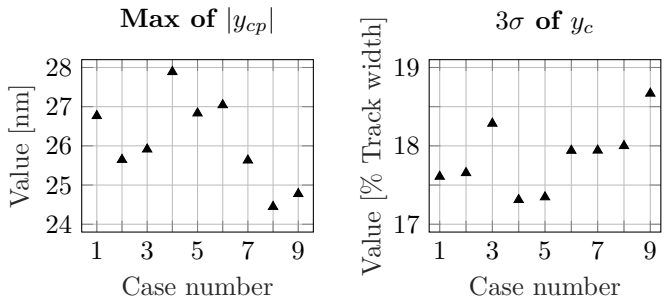

The closed loop system with the designed controller was simulated with uniformly distributed noises for 9 different plants to evaluate the performance and robustness of the controller. The total output of the DSA system and the output of the PZT actuator with the designed controller are plotted in Fig.~\ref{fig:ycycp}. The maximum values of the output of the PZT actuator and the $3\sigma$ values of the total output of the DSA system for all the 9 cases are shown in Fig. \ref{fig:maxycp3sigmayc}. It can be seen that the output of the PZT actuator remains within the desired limits for all the 9 cases. The worst case 3$\sigma$ value of the output of DSA was found to be 18.8\% of the the track width.   

\section{Conclusions}
\vspace{-10pt}
In this paper, a data driven control design framework has been used to design a robust track following controller for a benchmark dual stage actuator hard disk drive. 
The disturbance rejection problem has been formulated as a $H_2$ norm objective minimization problem with $H_\infty$ constraints for stability and shaping the closed loop transfer functions and $H_2$ norm constraints to limit the variances of signals. 
The performance of the controller was evaluated in simulation with uniformly distributed random noises and the results have been presented. 


\end{document}